\documentclass[a4paper,english]{jpconf}
\usepackage[T1]{fontenc}
\usepackage[latin1]{inputenc}

\makeatletter

\providecommand{\boldsymbol}[1]{\mbox{\boldmath $#1$}}

\providecommand{\tabularnewline}{\\}

\usepackage{bm}
\usepackage{slashed}
\usepackage{multirow}
\usepackage{threeparttable}

\newcommand{\ssst}[1]{\scriptscriptstyle{#1}}
\newcommand{\wt}[1]{\widetilde{#1}}
\newcommand{\mN}{m_{\ssst{N}}}

\usepackage{babel}
\makeatother
\begin{document}
\begin{flushright}

LA-UR-07-0730

\end{flushright}

\title{Strangeness-Conserving Hadronic Parity Violation at Low Energies}
\author{C.-P. Liu}
\address{T-16, B283, Theoretical Division, Los Alamos National Laboratory, Los Alamos, NM 87545, USA}
\ead{cpliu@lanl.gov}

\begin{abstract}
The parity-violating nucleon-nucleon interaction is the key to understanding
the strangeness-conserving hadronic weak interaction at low energies.
In this brief talk, I review the past accomplishments in and current
status of this subject, and outline a new joint effort between experiment
and theory that tries to address the potential problems in the past
by focusing on parity violation in few-nucleon systems and using the
language of effective field theory.
\end{abstract}

\section{Introduction}

Fifty years ago, the seminal paper on parity violation in weak interactions
by Lee and Yang, and the subsequent experimental confirmations in
$\beta$ decay of $^{60}\mbox{Co}$, muon, and pion is one of the
most exciting moments in physics. This discovery fully exemplifies
symmetry being a critical character of physical laws; through the
study of its conservation or violation, we are able to explore the
fundamental interactions and anything beyond our current knowledge
boundary. The fruitful achievements can be best summarized in the
very successful Standard Model of elementary particle physics, which
is based on the $SU(3)\otimes SU(2)\otimes U(1)$ gauge symmetry and
provides excellent descriptions of strong, weak, and electromagnetic
interactions.

The study of strangeness-conserving ($\Delta S=0$) hadronic weak
interaction, i.e., the weak interaction between two quarks without
change of flavor, was started shortly after the discovery of parity
violation~\cite{Tanner:1957}. However, not until a decade later
was the first evidence found by observing a non-zero circular polarization,
$P_{\gamma}=-(6\pm1)\times10^{-6}$, in the $\gamma$-decay of $^{181}\mbox{Ta}$~\cite{Lobashov:1967}.
The same Leningrad group then performed the first measurement of parity
violation in simple nuclear systems using radiative thermal neutron
capture by proton, and reported a result $P_{\gamma}=-(1.30\pm0.45)\times10^{-6}$~\cite{Lobashov:1972}
which surprised theoretical expectations not only by its being two
order of magnitude bigger but also by the sign. This famous Lobashov
experiment was eventually redone in the early 80s, and the result
$P_{\gamma}=(1.8\pm1.8)\times10^{-7}$~\cite{Knyaz'kov:1984}, though
with a big error, is now consistent with theory.

Despite lots of efforts being spent in this field, compared to what
have been achieved in leptonic, semi-leptonic, and strangeness-changing
hadronic weak interactions, our understanding in the $\Delta S=0$
hadronic sector is still relatively poor. The difficulties are not
hard to grasp: Since nucleon-nucleon ($NN$) systems are by far the
only viable venue to observe such an interaction,~%
\footnote{There are theoretical explorations of parity violation in processes
such as pion photo- and electro-productions (Refs.~\cite{Chen:2000hb,Chen:2000km}).
However, they have not been experimentally realized.%
} experimentally, one needs high precision to discern the much smaller
parity-violating (PV) signals. Theoretically, the non-perturbative
character of the quark-gluon dynamics makes a \char`\"{}first-principle\char`\"{}
formulation of the PV $NN$ interaction, which one needs to interpret
experiments, as yet impossible.

Given the fact that most weak interactions are tested so well, why
do we still bother with the $\Delta S=0$ sector? There are several
reasons for it. First: this is the only case where one can study the
neutral weak interaction between two quarks---the $\Delta S\neq0$
sector involves only charged currents. Therefore, we still need this
missing piece to make the whole weak interaction theory complete.
Second, and maybe more important from the modern perspective: as this
interaction comes out as an intricate interplay between the fundamental
weak interaction and the nonperturbative QCD, it can, in another way,
provide additional information about the low-energy strong dynamics,
which is not probed by usual scattering processes. Third, and somewhat
related to the previous one: we know, in the $\Delta S=1$ hadronic
weak interaction, the famous $\Delta I=1/2$ rule---a good example
of how strong interaction modifies the fundamental weak interaction.
It would be valuable to have some complementarity in the $\Delta S=0$
sector. Last but not least: several PV experiments via semi-leptonic
processes are in fact complicated by hadronic contributions. Two examples
which are particularly of interest to this meeting are i) the PV electron
proton or deuteron scattering which aims to explore the strangeness
content of a nucleon and ii) the atomic PV and Qweak experiments which
try to determine how the Weinberg angle $\sin^{2}\theta_{W}$ evolves
with $Q^{2}$ ($\sim0$ for the former and $\sim10^{-2}\,\textrm{GeV}^{2}$
for the latter).~%
\footnote{For these topics, please refer to contributions by D. Armstrong, K.
Paschke, and P. Souder in the same volume.%
} For better interpretation of these experiments, the hadronic contributions
appearing in terms of axial form factor or anapole moment should be
properly taken into account.

This short review is organized as following: The past accomplishments
in and the current status of the PV $NN$ interaction are first reviewed
in section~\ref{sec:old}. A new joint effort between experiment
and theory that tries to address the potential problems in the past
by focusing on parity violation in few-nucleon systems and using the
language of effective field theory (EFT) is then outlined in section~\ref{sec:new},
followed by a brief summary in section~\ref{sec:summary}.

\section{The Old Paradigm~\label{sec:old}}

\begin{table}

\caption{The five $S$--$P$ amplitudes, where $I$ denotes the isospin and
so do the superscripts in $v$'s and $\lambda$'s, and the entries
in the last column are the corresponding Danilov parameters.~\label{cap:S-Pamp}}

\vspace{0.2cm}

{\centering

\begin{tabular}{|c|c|c|c|c|c|c|c|}
\hline 
Transition&
$I\leftrightarrow I'$&
$\Delta I$&
$n$--$n$&
$n$--$p$&
$p$--$p$&
Amp.&
$E\rightarrow0$\tabularnewline
\hline 
$^{3}S_{1}\leftrightarrow^{1}P_{1}$&
$0\leftrightarrow0$&
0&
&
$\surd$&
&
$u$&
$\lambda_{t}$\tabularnewline
\hline 
&
&
0&
$\surd$&
$\surd$&
$\surd$&
$v^{0}$&
$\lambda_{s}^{0}$\tabularnewline
\cline{3-3} \cline{4-4} \cline{5-5} \cline{6-6} \cline{7-7} \cline{8-8} 
\multicolumn{1}{|c|}{$^{1}S_{0}\leftrightarrow^{3}P_{0}$}&
\multicolumn{1}{c|}{$1\leftrightarrow1$}&
1&
$\surd$&
&
$\surd$&
$v^{1}$&
$\lambda_{s}^{1}$\tabularnewline
\cline{3-3} \cline{4-4} \cline{5-5} \cline{6-6} \cline{7-7} \cline{8-8} 
\multicolumn{1}{|c|}{}&
\multicolumn{1}{c|}{}&
2&
$\surd$&
$\surd$&
$\surd$&
$v^{2}$&
$\lambda_{s}^{2}$\tabularnewline
\hline 
$^{3}S_{1}\leftrightarrow^{3}P_{1}$&
$0\leftrightarrow1$&
1&
&
$\surd$&
&
$w$&
$\rho_{t}$\tabularnewline
\hline
\end{tabular}

}
\end{table}

At low energies, two nucleon scattering mainly goes through the $S$-wave
channel, therefore, the PV $NN$ interaction, $V^{\ssst{\mathrm{PV}}}$,
then induces a small $P$-wave admixture. It is first pointed out
by Danilov~\cite{Danilov:1965,Danilov:1971fh,Danilov:1972} that,
at low energies, $V^{\ssst{\mathrm{PV}}}$ can be fully characterized
by five such $S\textrm{--}P$ scattering amplitudes as tabularized
in table~\ref{cap:S-Pamp}. Their zero-energy limits, the so-called
Danilov parameters, are the quantities to be determined phenomenologically.
As these two-body PV experiments were beyond experimental capability
in those times, most measurements were performed in heavier nuclei.
For this purpose, Desplanques and Missimer~\cite{Desplanques:1978mt}
extend this idea, by applying the Bethe-Goldstone equation, to many-body
systems. This PV potential in terms of the $S$--$P$ amplitudes takes
the form \begin{eqnarray}
V_{\ssst{S-P}}^{\ssst{\mathrm{PV}}}(\bm r) & = & \frac{4\,\pi}{\mN}\,\bigg\{\left[v^{0}\,\frac{1}{4}\,(3+\tau_{\cdot})+v^{1}\,\tau_{+}^{z}+v^{2}\,\tau^{zz}\right]\,\bm\sigma_{-}\cdot\{-i\,\bm\nabla\,,\,\delta(\bm r)\}\nonumber \\
 &  & +u\,\frac{1}{4}\,(1-\tau_{\cdot})\,\bm\sigma_{-}\cdot\{-i\,\bm\nabla\,,\,\delta(\bm r)\}+w\,\tau_{-}^{z}\,\bm\sigma_{+}\cdot\{-i\,\bm\nabla\,,\,\delta(\bm r)\}\bigg\}\,,\end{eqnarray}
 where $\mN$ is the nucleon mass, $\tau_{\cdot}\equiv\bm\tau_{1}\cdot\bm\tau_{2}$,
$\tau_{\pm}^{z}\equiv(\tau_{1}^{z}\pm\tau_{2}^{z})/2$, $\tau_{\times}^{z}\equiv i\,(\bm\tau_{1}\times\bm\tau_{2})^{z}/2$,
and $\tau^{zz}\equiv\left(3\,\tau_{1}^{z}\,\tau_{2}^{z}-\bm\tau_{1}\cdot\bm\tau_{2}\right)/(2\,\sqrt{6})$
are the isospin operators; $\bm\sigma_{\pm}\equiv\bm\sigma_{1}\pm\bm\sigma_{2}$
and $\bm\sigma_{\times}\equiv i\,\bm\sigma_{1}\times\bm\sigma_{2}$
are the spin operators.~%
\footnote{In the actual analyses by Desplanques and Missimer, the one-pion-exchange
contribution is added separately in order to better present the long-range
part of the $w$-type amplitude, i.e., the $^{3}S_{1}$--$^{3}P_{1}$
transition.~\label{SPana}%
} Though a lot of pre-80s data are analyzed in this framework~\cite{Desplanques:1978mt,Desplanques:1979ih,Desplanques:1980},
no detailed consistency check has been performed along this line;
and this framework has almost been forgotten after 80s, partly because
the meson-exchange picture gets more popularity.

The formulations of $V^{\ssst{\mathrm{PV}}}$ in terms of meson exchange,
which can be dated back to the works by Blin-Stoyle~\cite{Blin-Stoyle:1960a,Blin-Stoyle:1960b}
and Barton~\cite{Barton:1961eg}, are in fact not much younger than
the $S$--$P$ amplitude framework. Using the Barton's theorem---which
forbids any PV coupling between a nucleon and a neutral pseudoscalar
meson by $CP$ invariance---and considering mesons below GeV scale,
one is left with $\pi^{\pm}$, $\rho$, and $\omega$ mesons. The
resulting potential has seven PV meson-nucleon couplings, $h_{x}^{i}$'s
($x$ denotes the type of meson and $i$ the isospin) and takes the
form \begin{eqnarray}
V_{\ssst{\mathrm{OME}}}^{\ssst{\mathrm{PV}}}(\bm r) & = & V_{\rho,\omega}^{\ssst{\mathrm{PV}}}(\bm r)+V_{\pi}^{\ssst{\mathrm{PV}}}(\bm r)\,,\\
V_{\rho,\omega}^{\ssst{\mathrm{PV}}}(\bm r) & = & \frac{-1}{\mN}\,\Big\{ g_{\rho}\,\left[h_{\rho}^{0}\,\tau_{\cdot}+h_{\rho}^{1}\,\tau_{+}^{z}+h_{\rho}^{2}\,\tau^{zz}\right]\,\big(\bm\sigma_{-}\cdot\bm u_{\rho+}(\bm r)+i\,(1+\chi_{\rho})\,\bm\sigma_{\times}\cdot\bm u_{\rho-}(\bm r)\big)\nonumber \\
 &  & +g_{\omega}\,\left[h_{\omega}^{0}\,+h_{\omega}^{1}\,\tau_{+}^{z}\right]\,\big(\bm\sigma_{-}\cdot\bm u_{\omega+}(\bm r)+i\,(1+\chi_{\omega})\,\bm\sigma_{\times}\cdot\bm u_{\omega-}(\bm r)\big)\nonumber \\
 &  & -g_{\rho}\, h_{\rho}^{1}\,\tau_{-}^{z}\,\bm\sigma_{+}\cdot\bm u_{\rho+}(\bm r)+g_{\omega}\, h_{\omega}^{1}\,\tau_{-}^{z}\,\bm\sigma_{+}\cdot\bm u_{\omega+}(\bm r)\nonumber \\
 &  & +g_{\rho}\, h_{\rho}^{1'}\,\tau_{\times}^{z}\,\bm\sigma_{+}\cdot\bm u_{\rho-}(\bm r)-\Big\}\,,\\
V_{\pi^{\pm}}^{\ssst{\mathrm{PV}}}(\bm r) & = & \frac{1}{\sqrt{2}\,\mN}\, g_{\pi}\, h_{\pi}^{1}\,\tau_{\times}^{z}\,\bm\sigma_{+}\cdot\bm u_{\pi-}(\bm r)\,,\end{eqnarray}
where $g_{x}$'s denote the parity-conserving (PC) $x$-meson-nucleon
couplings; $\chi_{\omega}$ and $\chi_{\rho}$ are the isoscalar and
isovector strong tensor couplings, respectively; the spatial operator
$\bm u_{x-(+)}(\bm r)$ is defined as the (anti-) commutator of $-i\,\bm\nabla$
with the Yukawa function $f_{x}(r)$\begin{equation}
\bm u_{x\pm}(\bm r)=[-i\,\bm\nabla\,,\, f_{x}(r)]_{\pm}\equiv\left[-i\,\bm\nabla\,,\,\frac{\textrm{e}^{-m_{x}\, r}}{4\,\pi\, r}\right]_{\pm}\,.\end{equation}
 The attractiveness of this meson-exchange picture is apparent: not
only there are not much more undetermined parameters, but also it
provides a transparent gateway between phenomenology and the underlying
theory---one can actually perform hadronic calculations of these couplings
and compare with fits from experiments. This above form then becomes
the standard in this field after Desplanques, Donoghue, and Holstein
(DDH) give their prediction for these meson-nucleon coupling constants,
based on a quark model calculation~\cite{Desplanques:1980hn}.

Several hadronic calculations of these coupling constants are compared
in table~\ref{cap:pvcoupling}. The {}``best'' guess values by
DDH are pretty consistent with two other quark model calculations
(DZ~\cite{Dubovik:1986pj} and FCDH~\cite{Feldman:1991tj}). However,
as stressed by DDH, their best values have to be understood in the
context of the very modest {}``allowed'' ranges (the second column
of table~\ref{cap:pvcoupling}), which are a rough estimate of potentially
huge theoretical uncertainties. Thus, for example, despite the DDH
best value for $h_{\pi}^{1}$ differs by an order of magnitude from
the prediction of another quark model, the chiral soliton model (KM~\cite{Kaiser:1988bt,Kaiser:1989fd}),
these two results are still well within the allowed range. For the
most interested quantity $h_{\pi}^{1}$, there are two consistent
QCD Sum Rule calculations (HHK~\cite{Henley:1995ad,Henley:1998xh}
and Lobov~\cite{Lobov:2002xb}), which seem to favor DDH's result.
But, it is still premature to make any judgment at this stage, and
will be very interesting to see how the proposed lattice QCD effort
(BS~\cite{Beane:2002ca}), if ever realized, can provide us a more
definitive answer. 

\begin{table}

\caption{Hadronic predictions for the seven PV meson-nucleon coupling constants
(see text for explanation of abbreviations and references).~\label{cap:pvcoupling}}

\vspace{0.2cm}

{\centering

\begin{tabular}{rrcrrrrrrrr}
\hline 
&
\multicolumn{6}{c}{Quark Model}&
$\chi$-Soliton&
\multicolumn{2}{c}{QCD SR}&
LQCD\tabularnewline
$\times10^{7}$&
\multicolumn{3}{c}{DDH Range}&
Best&
DZ&
FCDH&
KM&
HHK&
Lobov&
BS\tabularnewline
\hline
\hline 
$h_{\pi}^{1}$&
0.0&
$\leftrightarrow$&
11.4&
4.6&
1.1&
2.7&
0.2&
3.0&
3.4&
proposed\tabularnewline
$h_{\rho}^{0}$&
-30.8&
$\leftrightarrow$&
11.4&
-11.4&
-8.4&
-3.8&
-3.7&
&
&
\tabularnewline
$h_{\rho}^{1}$&
-0.4&
$\leftrightarrow$&
0.0&
-0.2&
0.4&
-0.4&
-0.1&
&
&
\tabularnewline
$h_{\rho}^{2}$&
-11.0&
$\leftrightarrow$&
-7.6&
-9.5&
-6.8&
-6.8&
-3.3&
&
&
\tabularnewline
$h_{\omega}^{0}$&
-10.3&
$\leftrightarrow$&
5.7&
-1.9&
-3.8&
-4.9&
-6.2&
&
&
\tabularnewline
$h_{\omega}^{1}$&
-1.9&
$\leftrightarrow$&
-0.8&
-1.1&
-2.3&
-2.3&
-1.0&
&
&
\tabularnewline
$h_{\rho}^{1'}$&
&
&
&
0.0&
&
&
-2.2&
&
&
\tabularnewline
\hline
\end{tabular}

}
\end{table}

On the other hand, what are the experimental constraints on these
PV coupling constants? Although quite a few data has been accumulated
during the past years, not all of them have small errors to be constrictive.
Using the several precise data available to date, it is fair to say
that experiment and theory have not reached consistency. Two majors
puzzles are illustrated in FIG.~1 of Ref.~\cite{Haxton:2001mi}
and FIG.~8 of Ref.~\cite{Carlson:2001ma}. In the former figure,
a less ambitious two dimensional fit to some linear combinations of
these seven couplings is plotted. The constraints from the anapole
moments of $^{133}\textrm{Cs}$ and $^{205}\textrm{Tl}$ clearly contradict
with each other. Although the $^{133}\textrm{Cs}$ result probes the
similar linear combination of PV couplings as ones of the $\vec{p}\,\alpha$
and $^{19}\textrm{F}$ experiments, it favors larger values. If one
discards anapole constraints, the nuclear PV data do have an agreed
region, where the isoscalar coupling combination agrees with the DDH
best guess, but the isovector coupling combination, mostly determined
by $h_{\pi}^{1}$ (it is denoted as $f_{\pi}$ there), favors a much
smaller value than the DDH best guess---however, still in the allowed
range. The latter tight constraint, set by the $^{18}\textrm{F}$
data, is thought to be pretty robust as the experiments are repeated
by five different groups and the nuclear matrix elements are calibrated
well from analogous $\beta$ decay data. The second puzzle comes from
the $\vec{p}\, p$ asymmetry measurements performed at $13.6$, $45$,
and $221$ MeV. In the $V_{\ssst{\mathrm{OME}}}^{\ssst{\mathrm{PV}}}$
framework, these asymmetries only depend on two linear combinations
of the PV couplings: $h_{\rho}=h_{\rho}^{0}+h_{\rho}^{1}+h_{\rho}^{2}/\sqrt{6}$
and $h_{\omega}=h_{\omega}^{0}+h_{\omega}^{1}$; therefore, $h_{\rho}$
and $h_{\omega}$ should be uniquely determined by these experiments.
As one can see from the latter figure, the fitted value of $h_{\rho}\sim-20$
is consistent with the DDH best guess, but the one of $h_{\omega}\sim5$
is only marginally consistent with the very modest DDH reasonable
range and is in large discrepancy with other theoretical predictions.

\section{The New Direction~\label{sec:new}}

There can be many reasons for this unsettling situation: First, the
experiments might have their own problems. Second, as most data are
obtained in nuclear systems of medium to heavy mass, the reliability
of many-body calculations can be questioned. Third, one may wonder
if the meson-exchange model is really adequate.

In the last decade, we have seen quite some accomplishments in applying
the EFT technique to the construction of parity-conserving two- and
few-nucleon forces. Though there is still gap to catch up with the
success of modern, high-quality, phenomenological potentials, this
framework has the advantages of being completely general, model-independent,
and systematically improvable. Therefore, in order to avoid the potential
problems associated with the meson-exchange picture, Zhu et al. recently
applied the similar EFT technique and proposed a re-formulation of
$V^{\ssst{\mathrm{PV}}}$ to the order of $Q$ ($Q$ is the momentum
scale)~\cite{Zhu:2004vw}.

At this order, the PV potential, $V_{\ssst{\mathrm{EFT}}}^{\ssst{\mathrm{PV}}}$,
contains three components:~%
\footnote{An additional higher-order long-range term $V_{1,\mathrm{LR}}^{\mathrm{PV}}$
in Ref.~\cite{Zhu:2004vw} is omitted here, since it is shown to
be redundant~\cite{Liu:2006-u1}.%
} \begin{equation}
V_{\ssst{\mathrm{EFT}}}^{\ssst{\mathrm{PV}}}(\bm r)=V_{1,\ssst{\mathrm{{SR}}}}^{\mathrm{PV}}(\bm r)+V_{-1,\ssst{\mathrm{{LR}}}}^{\mathrm{PV}}(\bm r)+V_{1,\ssst{\mathrm{{MR}}}}^{\mathrm{PV}}(\bm r)\,.\end{equation}

1) The short-range (SR) part: This consists of the four-fermion contact
operators which meet all the symmetry requirements. It is expressed,
with 10 undetermined low-energy constants (LECs) $C$'s and $\wt{C}$'s,~%
\footnote{The notations here are of Ref.~\cite{Liu:2006dm}.%
} as \begin{eqnarray}
V_{1,\ssst{\mathrm{{SR}}}}^{\ssst{\mathrm{PV}}}(\bm r) & = & \frac{2}{\Lambda_{\chi}^{3}}\,\Big\{\left[C_{1}+(C_{2}+C_{4})\,\tau_{+}^{z}+C_{3}\,\tau_{\cdot}+C_{5}\,\tau^{zz}\right]\,\bm\sigma_{-}\cdot\bm y_{x-}(\bm r)\nonumber \\
 &  & +\left[\wt C_{1}+(\wt C_{2}+\wt C_{4})\,\tau_{+}^{z}+\wt C_{3}\,\tau_{\cdot}+\wt C_{5}\,\tau^{zz}\right]\,\bm\sigma_{\times}\cdot\bm y_{x-}(\bm r)\nonumber \\
 &  & +(C_{2}-C_{4})\,\tau_{-}^{z}\,\bm\sigma_{+}\cdot\bm y_{x+}(\bm r)+\wt C_{6}\,\tau_{\times}^{z}\,\bm\sigma_{+}\cdot\bm y_{x-}(\bm r)\Big\}\,,\end{eqnarray}
where $\Lambda_{\chi}$ is the scale of chiral symmetry breaking and
related to the pion decay constant $F_{\pi}$ by $\Lambda_{\chi}=4\,\pi\, F_{\pi}\approx1.161\,\textrm{GeV}$.
The spatial operator $\bm y_{m\pm}(\bm r)$ have the properties that
i) it is strongly peaked at $r=0$ with a range about $1/m_{x}$,
and ii) it approaches $\delta(\bm r)$ in the zero range (ZR) limit
(i.e., $m_{x}\rightarrow\infty$). A convenient choice, to mimic the
meson-exchange picture, is \begin{equation}
\bm y_{x\pm}(\bm r)=m_{x}^{2}\,\bm u_{x\pm}(\bm r)\rightarrow[-i\,\bm\nabla\,,\,\delta(r)/r^{2}]_{\pm}\,.\end{equation}
When one sets $m_{x}=m_{\rho,\omega}$ and assume the following relations
between $C$- and $\wt{C}$-type LECs

\begin{eqnarray}
\frac{\wt C_{1}}{C_{1}}=\frac{\wt C_{2}}{C_{2}} & = & 1+\chi_{\omega}\,,\\
\frac{\wt C_{3}}{C_{3}}=\frac{\wt C_{4}}{C_{4}}=\frac{\wt C_{5}}{C_{5}} & = & 1+\chi_{\rho}\,,\end{eqnarray}
then $V_{1,\ssst{\mathrm{{SR}}}}^{\ssst{\mathrm{PV}}}(\bm r)$ is
tantamount to the short-range sectors of $V_{\ssst{\mathrm{OME}}}^{\ssst{\mathrm{PV}}}(\bm r)$:
the ones corresponding the $\rho$- and $\omega$-exchanges $V_{\rho,\omega}^{\ssst{\mathrm{PV}}}(\bm r)$.
Also, in the ZR limit, one can see that $\bm y_{\infty-}(\bm r)$
and $\bm y_{\infty+}(\bm r)$ have the same matrix element. Therefore,
$V_{1,\ssst{\mathrm{{SR}}}}^{\ssst{\mathrm{PV}}}(\bm r)$ can be mapped
to $V_{\ssst{S-P}}^{\ssst{\mathrm{PV}}}(\bm r)$ so that the 10 LECs
at the superficial level can be reduced to 5, which corresponds to
the number of the physical $S$--$P$ amplitudes. 

2) The long-range (LR) part: This is the leading order term in EFT
(subscripted as {}``-1'') and corresponds to the familiar PV one-pion-exchange
potential, \begin{equation}
V_{-1,\ssst{\mathrm{{LR}}}}^{\ssst{\mathrm{PV}}}(\bm r)=V_{\pi^{\pm}}^{\ssst{\mathrm{PV}}}(\bm r)\,,\end{equation}
which depends on $h_{\pi}^{1}$.

3) The medium-range (MR) part: This is resulted from two-pion-exchange
(TPE) contributions with one of the four pion-nucleon couplings being
PV (therefore, also depends on $h_{\pi}^{1}$). It takes the form
\begin{eqnarray}
V_{1,\ssst{\mathrm{{MR}}}}^{\mathrm{PV}}(\bm r) & = & \frac{2}{\Lambda_{\chi}^{3}}\,\Bigg\{-4\,\sqrt{2}\,\pi\, g_{A}^{3}\, h_{\pi}^{1}\,\bm\sigma_{\times}\cdot\bm y_{2\pi}^{L}(\bm r)\nonumber \\
 &  & +3\,\sqrt{2}\,\pi\, g_{A}^{3}\, h_{\pi}^{1}\,\,\tau_{\times}^{z}\,\bm\sigma_{+}\cdot\left[\left(1-\frac{1}{3\, g_{A}^{2}}\right)\,\bm y_{2\pi}^{L}(\bm r)-\frac{1}{3}\,\bm y_{2\pi}^{H}(\bm r)\right]\Bigg\}\,,\\
\bm y_{2\pi}^{L}(\bm r) & = & \left[-i\,\bm\nabla\,,\,\mathrm{F.T.}\left(\frac{\sqrt{4\, m_{\pi}^{2}+\bm q^{2}}}{|\bm q|}\,\ln\left(\frac{\sqrt{4\, m_{\pi}^{2}+\bm q^{2}}+|\bm q|}{2\, m_{\pi}}\right)\right)\right]\,,\label{eq:y2piL}\\
\bm y_{2\pi}^{H}(\bm r) & = & \left[-i\,\bm\nabla\,,\,\mathrm{F.T.}\left(\frac{4\, m_{\pi}^{2}}{|\bm q|\,\sqrt{4\, m_{\pi}^{2}+\bm q^{2}}}\,\ln\left(\frac{\sqrt{4\, m_{\pi}^{2}+\bm q^{2}}+|\bm q|}{2\, m_{\pi}}\right)\right)\right]\,,\label{eq:y2piH}\end{eqnarray}
where {}``F.T.'' denotes a Fourier transform from $q$- to $r$-space.~%
\footnote{Some mistakes in Eq.~(121) of Ref.~\cite{Zhu:2004vw} have been
fixed in order to produce Eqs. (\ref{eq:y2piL}, \ref{eq:y2piH});
see Refs.~\cite{Hyun:2006mp,Liu:2006dm}.%
} Note that both the MR and SR interactions appear at the same EFT
order (next-to-next-to-leading, subscripted as {}``1''), and their
expressions should be understood in the context of the specific regularization
scheme. The MR interaction as given by Zhu et al. only contains the
non-analytic part of TPE; all the analytic part has been effectively
included in the SR interaction~\cite{Zhu:2004vw}.

Overall, this new formulation contains 6 (5 LECs plus $h_{\pi}^{1}$)
undetermined parameters. Though this number seems comparable to 7
in the OME framework, one has to note that this EFT formulation is
only to $O(Q)$; therefore, one should be very careful when trying
to analyse not-so-low-energy processes (e.g., $\vec{p}\, p$ scattering
at 221 MeV is obviously out of scope). On the other hand, if one further
limits the momentum scale under the pion mass, i.e., $Q\ll m_{\pi}\cong140\,\textrm{MeV}$,
which corresponds to an energy scale of $E\ll10\,\textrm{MeV}$, then
the pion degrees of freedom can be integrated out and this leads to
a pionless theory where only 5 LECs are needed. Although this pionless
framework requires less parameters (by one), the most interested PV
constant $h_{\pi}^{1}$ becomes obscure since it is implicitly included
in LECs.

For determining these six parameters, a search program has also been
sketched out in Ref.~\cite{Zhu:2004vw}. The basic idea is to explore
as many low-energy observables in few-nucleon systems as possible.
With the advance of experimental apparatus and techniques, PV experiments
in few nucleon systems with $10\%$ precision become feasible nowadays.
Furthermore, modern few-body calculations are also sophisticated enough
to allow reliable interpretations of these experiments. Combining
the model-independent PV $NN$ interaction based on EFT, we should
be able to properly address the above-mentioned problems that possibly
undermine a consistent picture of nuclear parity violation.

\begin{table}
\begin{center}

\begin{threeparttable}

\caption{The nuclear PV search program in few-nucleon systems (non-exhaustive).~\label{cap:fbPV}}

\begin{tabular}{|l|l|l|}
\hline 
Observables&
Theory&
Experiment $(\times10^{7})$\tabularnewline
\hline
\hline 
$A_{L}^{\vec{p}p}(13.6\,$MeV)&
$-0.45\,\mN\lambda_{s}^{pp}$~\tnote{a}&
$-0.93\pm0.21$ (Bonn)~\cite{Eversheim:1991tg}\tabularnewline
$A_{L}^{\vec{p}p}(45\,$MeV)&
$-0.78\,\mN\lambda_{s}^{pp}$~\tnote{a}&
$-1.57\pm0.23$ (SIN)~\cite{Kistryn:1987tq}\tabularnewline
\hline 
$\frac{d}{d\, z}\,\phi_{n}^{\vec{n}p}(\textrm{th.})$&
$\left[2.50\,\lambda_{s}^{np}-0.57\,\lambda_{t}+1.41\,\rho_{t}\right]\,\mN+0.29\,\wt C_{6}^{\pi}$~\tnote{a}&
SNS\tabularnewline
\hline 
$P_{\gamma}^{np}($th.)&
$\left[-0.16\,\lambda_{s}^{np}+0.67\,\lambda_{t}\right]\,\mN$~\tnote{a}&
$(1.8\pm1.8)$~\cite{Knyaz'kov:1984}, SNS?\tabularnewline
$A_{L}^{\vec{\gamma}d}(1.32\,$keV)&
Same as above~\tnote{a}&
HIGS? IASA? Spring-8?\tabularnewline
\hline 
$A_{\gamma}^{\vec{n}p}($th.)&
$-0.09\,\mN\rho_{t}-0.27\,\wt C_{6}^{\pi}$~\tnote{a}&
$(0.6\pm2.1)$~\cite{Cavaignac:1977uk}, LANSCE~\tnote{c}\tabularnewline
\hline 
$\frac{d}{d\, z}\,\phi_{n}^{\vec{n}d}($th.)&
To be done&
SNS?\tabularnewline
\hline 
$A_{\gamma}^{\vec{n}d}($th.)&
$\left[0.59\,\lambda_{s}^{nn}+0.51\,\lambda_{s}^{np}+1.18\,\lambda_{t}+1.42\,\rho_{t}\right]\,\mN$~\tnote{b}&
$(42\pm38)$~\cite{Alberi:1988fd}, SNS?\tabularnewline
\hline 
$A_{L}^{\vec{p}\alpha}(46\,$MeV)&
$\left[-0.48\,\lambda_{s}^{pp}-0.24\,\lambda_{s}^{np}-0.54\,\lambda_{t}-1.07\,\rho_{t}\right]\,\mN$~\tnote{b}&
$-3.3\pm0.9$ (SIN)~\cite{Lang:1985jv}\tabularnewline
\hline 
$\frac{d}{d\, z}\,\phi_{n}^{\vec{n}\alpha}($th.)&
$\left[1.2\,\lambda_{s}^{nn}+0.6\,\lambda_{s}^{np}+1.34\,\lambda_{t}-2.68\,\rho_{t}\right]\,\mN$~\tnote{b}&
$(8\pm14)$~\cite{Markoff:1997}, NIST~\tnote{c}\tabularnewline
\hline
\end{tabular}

\begin{tablenotes}

\item[a)] Results taken from Ref.\,\cite{Liu:2006dm}.

\item[b)] Results taken from Ref.\,\cite{Zhu:2004vw}. These calculations have to be improved, and also be checked with calculations using the pionful theory since it is questionable if the pionless framework can apply to these cases (See Ref.\,\cite{Liu:2006dm} for some remarks).

\item[c)] With plans continuing at SNS.

\end{tablenotes}

\end{threeparttable}

\end{center}
\end{table}

Both experimental and theoretical efforts in this broad program are
summarized, but not exhaustively, in table~\ref{cap:fbPV}. 

On the experimental side, there are two existing data points, the
low-energy $\vec{p}\, p$ scattering (the 13.6 and 45 MeV experiments
measure virtually the same parameter) and $\vec{p}\,\alpha$ scattering.
There are two ongoing experiments: the asymmetry measurement in radiative
polarized neutron capture on proton at the Los Alamos Neutron Science
Center (LANSCE) and the thermal neutron spin rotation measurement
in liquid helium at the National Institute of Standard and Technology
(NIST). The Fundamental Neutron Physics Beamline (FNPB) program, which
is going to take advantage of the high-intensity, pulsed neutron beam
from the just-completed Spallation Neutron Source (SNS), will consider
other neutron-induced processes. Overall, it looks very promising
that enough data will be taken in the near future.

On the theoretical side, the re-analysis of these PV observables is
also under way~\cite{Liu:2006dm}. This is done in the so-called
{}``hybrid'' EFT fashion, which marries the general PV potential
derived from the EFT consideration and the start-of-the-art nuclear
wave functions obtained from phenomenological model calculations.
The 5 independent LECs are completely mapped to the dimensionless
Danilov parameters: $m_{N}\times(\lambda_{s}^{pp},\,\lambda_{s}^{nn},\,\lambda_{s}^{np},\,\lambda_{t},\,\rho_{t})$
with the long-range one-pion-exchange contribution, characterized
by $\wt C_{6}^{\pi}\propto h_{\pi}^{1}$, being singled out from the
$^{3}S_{1}$--$^{3}P_{1}$ amplitude.~%
\footnote{In this sense, this new framework is a revival of the $S$--$P$ analysis
proposed by Danilov, Desplanques and Missimer (also see footnote~\ref{SPana}).%
} The observables in two-body systems have been analyzed with the results
given in the table~\ref{cap:fbPV}. The observables in few-body systems
should be analyzed in the same way with updated calculations. These
results will be valuable for prioritizing the future measurements.

\section{Summary~\label{sec:summary}}

The study of strangeness-conserving hadronic weak interaction is a
challenging task both experimentally and theoretically. Although the
efforts in the past fifty years have not been able to provide us a
consistent overall picture, the precious lessons learned however motivate
a new and promising direction. This new research program consists
of three important ingredients: (1) the high-precision measurements
of nuclear parity-violation in few-nucleons systems, (2) the reliable
few-body calculations using the state-of-the-art techniques to interpret
the experiments, and (3) the general, model-independent formulation
of the parity-violating nucleon-nucleon interaction, which in combination
aim to address the potential problems in the past. The Fundamental
Neutron Physics Beamline program at the Spallation Neutron Source
is going to trigger a new renaissance for this research, and with
intensive joint efforts between experiment and theory, one hopes not
only to reach a consistent picture of hadronic weak interaction but
also to provide important, additional input for the study of the nonperturbative
dynamics of strong interaction.

\newpage

\section*{References}

\bibliographystyle{iopart-num}
\bibliography{GHP06_Liu}

\providecommand{\newblock}{}
\begin{thebibliography}{10}
\expandafter\ifx\csname url\endcsname\relax
  \def\url#1{{\tt #1}}\fi
\expandafter\ifx\csname urlprefix\endcsname\relax\def\urlprefix{URL }\fi
\providecommand{\eprint}[2][]{\url{#2}}

\bibitem{Tanner:1957}
Tanner N 1957 {\em Phys. Rev.\/} {\bf 107} 1203--1204

\bibitem{Lobashov:1967}
Lobashov V~M, Nazarenko V~A, Saenko L~F, Smotritsky L~M and Kharkevitch G~I
  1967 {\em Phys. Lett.\/} {\bf 25B} 104--106

\bibitem{Lobashov:1972}
Lobashov V~M {\em et~al.\/} 1972 {\em Nucl. Phys.\/} {\bf A197} 241--258

\bibitem{Knyaz'kov:1984}
Knyaz'kov V~A {\em et~al.\/} 1984 {\em Nucl. Phys.\/} {\bf A417} 209--230

\bibitem{Chen:2000hb}
Chen J~W and Ji X~D 2001 {\em Phys. Rev. Lett.\/} {\bf 86} 4239--4242

\bibitem{Chen:2000km}
Chen J~W and Ji X~D 2001 {\em Phys. Lett. B\/} {\bf 501} 209--215

\bibitem{Danilov:1965}
Danilov G~S 1965 {\em Phys. Lett.\/} {\bf 18} 40--41

\bibitem{Danilov:1971fh}
Danilov G~S 1971 {\em Phys. Lett.\/} {\bf 35B} 579--580

\bibitem{Danilov:1972}
Danilov G~S 1972 {\em Sov. J. Nucl. Phys.\/} {\bf 14} 443

\bibitem{Desplanques:1978mt}
Desplanques B and Missimer J 1978 {\em Nucl. Phys.\/} {\bf A300} 286--312

\bibitem{Desplanques:1979ih}
Desplanques B and Missimer J 1979 {\em Phys. Lett.\/} {\bf 84B} 363--367

\bibitem{Desplanques:1980}
Desplanques B 1980 {\em Nucl. Phys.\/} {\bf A335} 147--167

\bibitem{Blin-Stoyle:1960a}
Blin-Stoyle R~J 1960 {\em Phys. Rev.\/} {\bf 118} 1605--1607

\bibitem{Blin-Stoyle:1960b}
Blin-Stoyle R~J 1960 {\em Phys. Rev.\/} {\bf 120} 181--189

\bibitem{Barton:1961eg}
Barton G 1961 {\em Nuovo Cimento\/} {\bf 19} 512--527

\bibitem{Desplanques:1980hn}
Desplanques B, Donoghue J~F and Holstein B~R 1980 {\em Ann. Phys. (N.Y.)\/}
  {\bf 124} 449--495

\bibitem{Dubovik:1986pj}
Dubovik V~M and Zenkin S~V 1986 {\em Ann. Phys.\/} {\bf 172} 100--135

\bibitem{Feldman:1991tj}
Feldman G~B, Crawford G~A, Dubach J and Holstein B~R 1991 {\em Phys. Rev. C\/}
  {\bf 43} 863--874

\bibitem{Kaiser:1988bt}
Kaiser N and Meissner U~G 1988 {\em Nucl. Phys.\/} {\bf A489} 671--682

\bibitem{Kaiser:1989fd}
Kaiser N and Meissner U~G 1989 {\em Nucl. Phys.\/} {\bf A499} 699--726

\bibitem{Henley:1995ad}
Henley E~M, Hwang W~Y~P and Kisslinger L~S 1996 {\em Phys. Lett. B\/} {\bf 367}
  21--27

\bibitem{Henley:1998xh}
Henley E~M, Hwang W~Y~P and Kisslinger L~S 1998 {\em Phys. Lett. B\/} {\bf 440}
  449--450

\bibitem{Lobov:2002xb}
Lobov G~A 2002 {\em Phys. Atom. Nucl.\/} {\bf 65} 534--538

\bibitem{Beane:2002ca}
Beane S~R and Savage M~J 2002 {\em Nucl. Phys. B\/} {\bf 636} 291--304

\bibitem{Haxton:2001mi}
Haxton W~C, Liu C~P and Ramsey-Musolf M~J 2001 {\em Phys. Rev. Lett.\/} {\bf
  86} 5247--5250

\bibitem{Carlson:2001ma}
Carlson J, Schiavilla R, Brown V~R and Gibson B~F 2002 {\em Phys. Rev. C\/}
  {\bf 65} 035502

\bibitem{Zhu:2004vw}
Zhu S~L, Maekawa C~M, Holstein B~R, Ramsey-Musolf M~J and van Kolck U 2005 {\em
  Nucl. Phys. A\/} {\bf 748} 435--498

\bibitem{Liu:2006-u1}
Liu C~P and Ramsey-Musolf M~J In preparation

\bibitem{Liu:2006dm}
Liu C~P 2006  (\textit{Preprint} \eprint{nucl-th/0609078})

\bibitem{Hyun:2006mp}
Hyun C~H, Ando S and Desplanques B (\textit{Preprint} \eprint{nucl-th/0609015})

\bibitem{Eversheim:1991tg}
Eversheim P~D {\em et~al.\/} 1991 {\em Phys. Lett. B\/} {\bf 256} 11--14

\bibitem{Kistryn:1987tq}
Kistryn S {\em et~al.\/} 1987 {\em Phys. Rev. Lett.\/} {\bf 58} 1616--1619

\bibitem{Cavaignac:1977uk}
Cavaignac J~F, Vignon B and Wilson R 1977 {\em Phys. Lett.\/} {\bf 67B}
  148--150

\bibitem{Alberi:1988fd}
Alberi J {\em et~al.\/} 1988 {\em Can. J. Phys.\/} {\bf 66} 542--547

\bibitem{Lang:1985jv}
Lang J {\em et~al.\/} 1985 {\em Phys. Rev. Lett.\/} {\bf 54} 170--173

\bibitem{Markoff:1997}
Markoff D 1997 {\em Measurement of the Parity Nonconserving Spin-Rotation of
  Transmitted Cold Neutrons Through a Liquid Helium Target\/} Ph.D. thesis
  University of Washington

\end{thebibliography}

\end{document}